\documentclass[ prx, superscriptaddress, aps]{revtex4-2}
\usepackage{bm}
\usepackage{setspace}
\usepackage{graphicx}
\usepackage[margin=1in]{geometry}
\usepackage{amsmath, physics}
\usepackage{amsfonts}
\usepackage[sort&compress]{natbib}
\usepackage{notes2bib}
\usepackage{xcolor}
\begin{document}

\preprint{APS/123-QED}

\title{Transfer tensor analysis of localization in the Anderson and Aubry-André-Harper models}

\author{Michelle C. Anderson}
\affiliation{Department of Chemistry and Biochemistry, University of Nevada, Las Vegas, Las Vegas, Nevada 89154, USA}

\author{Chern Chuang}
\email{Contact author: chern.chuang@unlv.edu}
\affiliation{Department of Chemistry and Biochemistry, University of Nevada, Las Vegas, Las Vegas, Nevada 89154, USA}

\begin{abstract}
We use the transfer tensor method to analyze localization and transport in simple disordered systems, specifically the Anderson and Aubry-André-Harper models. Emphasis is placed on the memory effects that emerge when ensemble-averaging over disorder, even when individual trajectories are strictly Markovian. We find that transfer tensor memory effects arise to remove fictitious terms that would correspond to redrawing static disorder at each time step, which would create a temporally uncorrelated dynamic disorder. Our results show that while eternal memory is a necessary condition for localization, it is not sufficient. We determine that signatures of localization and transport can be found within the transfer tensors themselves by defining a metric called ``outgoing-pseudoflux''. This work establishes connections between theoretical research on dynamical maps and Markovianity and localization phenomena in physically realizable model systems.

\end{abstract}

\maketitle

\section{Introduction}

In photovoltaic applications, fast transport of excitons through crystal lattices is critical for engineering success\cite{akel2023,wang2024,blach2025} whereas in quantum information applications, particularly quantum memory, localization is equally important.\cite{chen2023,brown2016,chandrashekar2015,wootton2011} Dynamic localization phenomena originate from a periodic force,\cite{dunlap1986,tang2022,borovkova2015} but Anderson localization originates from destructive interference due to static disorder in the system.\cite{anderson1958} In some systems all energy eigenstates are localized whereas other systems display mobility edges, energies below which states are localized and above which they are delocalized.\cite{an2021,dwiputra2021environment} 

Theoretical study of quantum transport in realistic systems is complicated by the unfavorable scaling of computational requirements when modeling quantum systems,\cite{breuer2002} so simple but powerful treatments such as the Anderson,\cite{anderson1958}, Haken-Strobl-Reineker (HSR),\cite{reineker1982,haken1973} or Aubry-André-Harper (AAH) models\cite{aubry1980,harper1955} are often employed to understand transport and localization in tight binding lattices.  As many experimental measurements sample over regions with defects, inducing effects such as inhomogeneous broadening on spectroscopic measurements,\cite{kahle2022,blach2025,raimondo_2006,barvik1999,sun2023} 
it is common practice to average over static disorder in model parameters in order to better describe experimental observations.\cite{anderson1958,longhi2023,moix2013,arvi1999} 
These simple models have increased our understanding of phenomena including metal-insulator transitions and diffusion rates in the presence of thermal noise and disorder.\cite{mott1961,vollhardt1980,tarquini2017,longhi2023,dominguez2019,aubry1980,harper1955,moix2013}

The dynamics resulting from averaging over a static disorder ensemble recovers the spatial homogeneity that was destroyed when the disorder was introduced. However, a somewhat counterintuitive complication arises.
Evolutions consisting of mixtures of distinct unitary evolutions, averages over static disorder in our case, may show memory effects. Although the exact interpretation of these effects, and of quantum non-Markovianity in general, remains a matter of discussion,\cite{breuer2018,jagadish2020,budini2018,banacki2023,megier2017} the association of this apparent non-Markovianity and Anderson localization has been noted in previous studies.\cite{lorenzo2017,lorenzo2017_2,kropf2020} We will reserve the connection of this phenomenon to other established measures of quantum non-Markovianity and its potential connection to information backflow for future studies,\cite{Rivas2014,Breuer2016,breuer2018} and rather analyze the origin of these memory effects using transfer tensors (TT), which are ideally adapted to this purpose. Transfer tensors or related process tensors have proved useful characterizing environmental noise,\cite{pollock2018,chen2020,chen2022,chen2023} extracting information about emitted photons,\cite{cygorek2025} and making arguments about the divisibility of dynamical maps characterizing system evolution.\cite{gherardini2022} Here we restrict ourselves to stating that a system has memory when memory terms appear in the transfer tensor hierarchy without any intended implication about non-Markovianity, such as that arises from genuine non-perturbative system-bath interactions. \footnote{We note that as discussed in Ref.~\cite{gherardini2022}, having vanishing $k>1$-step tensors is a sufficient condition for CP-divisible evolutions and the dynamics is consequently Markovian by that criterion. Here, the TTs from mixtures of Markovian dynamics are (uniquely) constructed identically to that used in Ref.~\cite{gherardini2022}. However, there is no system-environment correlation in the traditional open quantum system sense } We see that memory terms differentiate static disorder from dynamic disorder and find that, although the existence of memory does not necessarily lead to localization, another characteristic of the transfer tensors, which we call \textit{outgoing-pseudoflux}, provides an indication of localization. This demonstrates the potential of TT as a diagnostic tool for assessing transport properties and understanding the origin and impact of memory derived from disorder averaging.

\section{Transfer Tensor Analysis of the Anderson Model}

We start from the 1D Anderson model, 
\cite{anderson1958,moix2013}
\begin{equation}
    H = \sum_n \epsilon_n |n \rangle \langle n| + V \sum_n |n-1 \rangle \langle n| + |n \rangle \langle n - 1|
    \label{eqham}
\end{equation}
in the lattice site basis, $|n\rangle$, which includes nearest-neighbor interactions with coupling strength $V$ and site energies, $\epsilon_n$, drawn from a normal distribution with zero mean and standard deviation $\sigma$. Throughout this paper, we set $\hbar=1$ and $V=1$. 
Introducing static disorder localizes the system in 1D and breaks spatial homogeneity,\cite{anderson1958,reineker1992} but averaging over many realizations of the static disorder restores it. Whereas the individual trajectories in the ensemble are Markovian and have no memory, the spatially homogeneous ensemble evolution now has memory, as noted in the context of convex mixing of quantum channels or dynamical maps\cite{breuer2002,breuer2018,jagadish2020} and in previous studies of Anderson localization.\cite{lorenzo2017,lorenzo2017_2} 
To better see this, we adopt the transfer tensor method to describe the disorder-averaged dynamics.

Fixing a time step $\Delta t$, a system at $t = k\Delta t$ given by the density matrix $\rho(k)$ can be described by a dynamical map, $M(k)$, such that $\rho(k) = M(k) \rho(0)$. Presuming time-translational invariance, the TT at time step $k$ is\cite{cerrillo2014,wu2024} 
\begin{equation}T(k) = M(k) - \sum_{l = 1}^{k-1} T(k-l) M(l )
\label{eq:ttm_rec}
\end{equation}
so that the density matrix after $k$ time steps is
\begin{equation}
    \rho(k) = \sum_{l=1}^{k} T(l)\rho(k-l).
    \label{Eq:tt_prop}
\end{equation}
The tensor $T(k)$ describes the effect on current dynamics traced to the system state $k$ time steps in the past. In the case where the system has a finite memory time, the TT becomes negligibly small at a certain $k=k_c$ and the series can be truncated while maintaining numerical accuracy when applying Eq.~(\ref{Eq:tt_prop}). In all cases, $T(1)=M(1)$, and if there is no memory, $T(2)$ and all subsequent tensors vanish. 

We will find that individual elements of the transfer tensor can provide useful information. Because we average over disorder and restore spatial homogeneity, we can analyze only the transfer tensor elements involving the center site of the system.  
The element $T_{m,n,o,p}(l)$ quantifies the influence that $\rho_{o,p}(k)$ will have on $\rho_{m,n}(k+l)$ and $T_{m,n,o,p}(l)$ depends only on the differences $m-p$, $n-p$, and $o-p$. The utility of an equivalent relationship for the generalized quantum master equations was recently recognized and exploited for performance enhancement.
\cite{bhattacharyya2024} The spatial homogeneity of the TT allows analysis with modest basis sizes.
\begin{figure}[h]
\includegraphics[width=14cm]{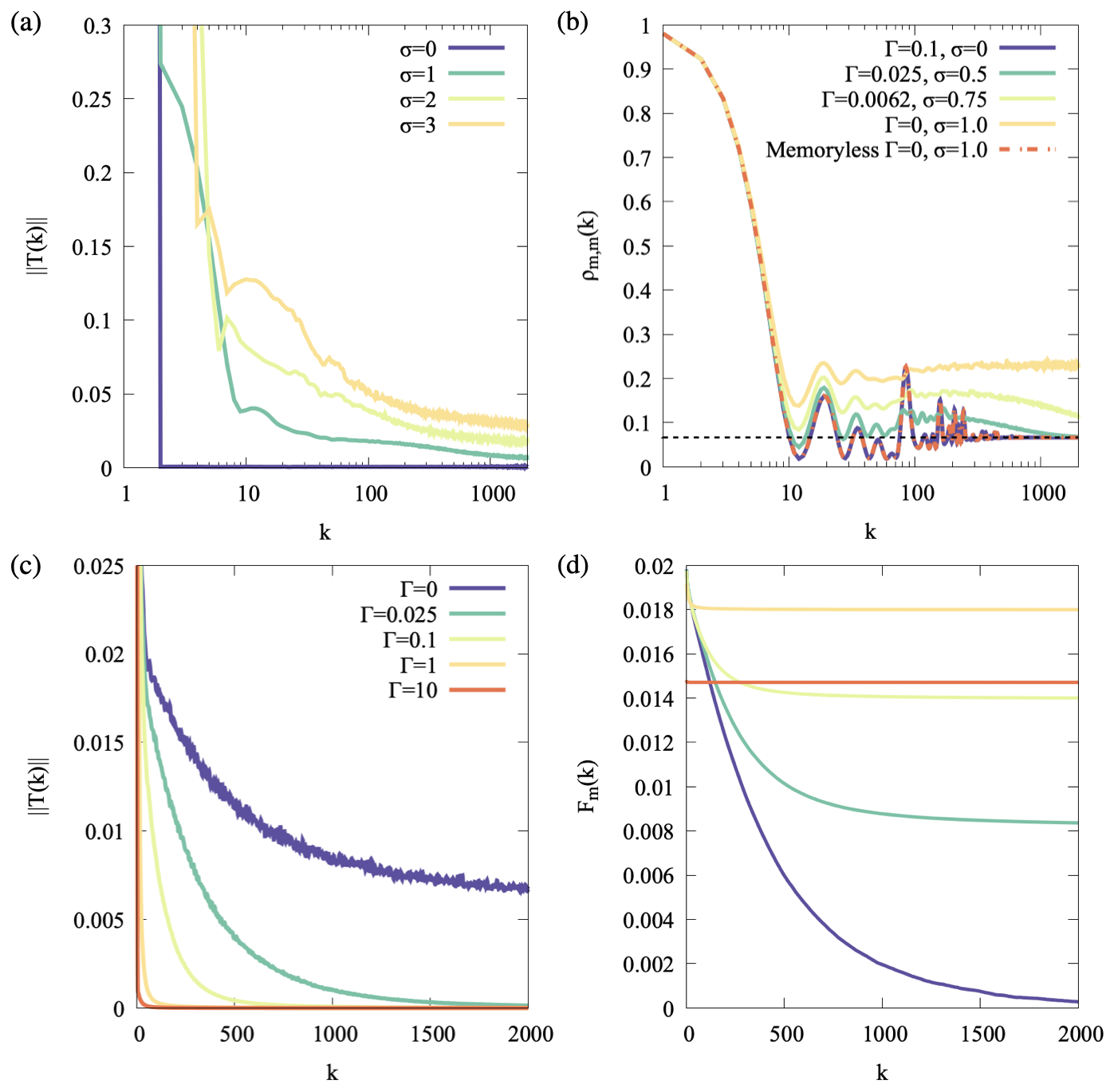}
\caption{  (a) The Frobenius norm of the transfer tensor $||T(k)||$ as a function of $k$ for  $\Gamma=0$ with different $\sigma$ values.  (b) Population of site $m=8$ following initialization in a delta function localized on $m$ as a function of time step for several ensembles including a case in which all transfer tensor memory terms were artificially set to zero. The dashed black line is the equilibrium population. (c) The Frobenius norm of the transfer tensor $||T(k)||$ as a function of time step $k$ for several different $\Gamma$ values with $\sigma=V$. (d) The cumulative outgoing-pseudoflux as a function of time step for the same $\Gamma$  and $\sigma$ values as in (c) showing convergence to non-zero values at finite $\Gamma$ and convergence to zero at  $\Gamma=0$. Note that, when $\Gamma=10V$, noise overpowers disorder and the system is effectively memoryless for the chosen $\Delta t$. All calculations in this figure involved a basis of size $15$, $V\Delta t = 0.1$ and $2000$ trajectories. In (a) and (c) the norm of the first transfer tensor is on the order of the basis size and not shown on these axes.}
\centering
\label{fig1}
\end{figure}

The memory that arises for the disorder-averaged ensembles is quantified in Fig. \ref{fig1} (a) which shows the Frobenius norms of the TT, $||T(k)||$, as a function of $k$ for several $\sigma$. In the case where $\sigma=0$, memory is zero within the numerical tolerance for all $k>1$. For $\sigma>0$, the tensor norms converge toward a finite value as $k \to \infty$. This finite value increases with $\sigma$ and suggests an \textit{eternal} memory, in the sense that there will be an indefinite number of memory terms in the transfer tensor. A similar phenomenon has been noted in the study of the eternally non-Markovian quantum master equation, which is characterized by a quasi-Gorini-Kossakowski-Sudarshan-Lindblad\cite{lindblad1976,gorini1976} form in which there is a perpetually negative rate associated with a jump operator, which is taken as an indication of quantum non-Markovianity.\cite{megier2017} It is important, however, to point out that the ensemble evolutions we address can be viewed as convex mixtures of Markovian maps. There are competing definitions of quantum non-Markovianity and there has been debate as to whether convex combinations of Markovian maps should be considered to demonstrate it.\cite{kropf2016,megier2017,budini2018,breuer2018,li2019,jagadish2020,kropf2020,budini2023} 
Specifically, information backflow as a measure of non-Markovianity has been debated in this context.\cite{megier2017,breuer2018,banacki2023} Whether the ensemble averaged systems we analyze should be considered to display quantum non-Markovianity or whether the appearance of memory terms in the transfer tensor in general is indicative of quantum non-Markovianity is beyond the current scope and we will avoid using this term in favor of discussing memory. When we refer to memory in this work, we refer only to the presence of transfer tensor memory terms, that the Frobenius norm of any $k>1$-step TT is finite.

The source of eternal memory in Fig. \ref{fig1} (a) is easily understood. Because the evolution of individual trajectories is Markovian, we can define the dynamical map for the disorder-averaged ensemble at any time step by $M(k ) = \frac{1}{N} \sum_{i=1}^N \left[ M^{(i)}(1) \right]^k$, where $N$ is the number of disordered trajectories and $M^{(i)}(1)$ represents the map of the $i$th trajectory over one time step. Using Eq.~(\ref{eq:ttm_rec}), we find
\begin{eqnarray}
     T(2) &=& \frac{1}{N} \sum_i \left[M^{(i)}(1)\right]^2 - \frac{1}{N^2} \sum_{i,j} M^{(i)}(1)M^{(j)}(1)
\end{eqnarray}
and see that ${T}(2) \ne 0$. This first memory term includes the dynamical map ${M}(2) $ necessary to propagate from time $0$ to $2\Delta t$, as well as a term which subtracts fictitious cross terms between individual trajectory's dynamical maps, $M^{(i)}M^{(j)}$, that are produced when $ M(1)$ is squared. These eliminated fictitious terms correspond to redrawing the randomly selected energies of the sites at each time step, or, equivalently, randomly selecting $H$ at each step, a process that would eliminate static disorder and substitute dynamic disorder. Similar fictitious terms must be eliminated for all $k>1$. This suggests that the role of memory terms is to differentiate the static and corresponding dynamic disorder, elaborated below. 

Upon closer examination, the Markovian evolution represented by $M(k)=[T(1)]^k$ in the above discussion is equivalent to the dynamics induced by averaging uncorrelated Gaussian white noise in the limit of $\Delta t \rightarrow  0$. This is actually the \textit{HSR model, averaged over stochastic noise},\cite{haken1973,reineker1982} 
where the site energies follow zero mean Gaussian statistics with no spatial or temporal correlations
 \begin{eqnarray}
     \langle\epsilon_n(t)\rangle&=&0,\\
     \langle\epsilon_n(t)\epsilon_m(t')\rangle&=&\Gamma\delta_{nm}\delta(t-t').
 \end{eqnarray} 
 This gives rise to the equivalent Gorini–Kossakowski–Sudarshan–Lindblad equation of motion for the corresponding noise-averaged density matrix,\cite{lindblad1976,gorini1976}
\begin{equation}
    \frac{d \rho(t)}{d t} = -i\left[ H, \rho(t) \right] - \frac{\Gamma}{2} \sum_n \left[ |n\rangle \langle n|, \left[ |n\rangle \langle n|, \rho(t) \right]\right].
    \label{eqn2}
\end{equation}
This connection immediately puts the role of $T(k>1)$ into the context of transport, since the HSR model has been well characterized to give rise to steady-state diffusive transport,\cite{haken1973,reineker1982,madhukar1977,rebentrost2008,kunsel2021} 
suggesting that the memory generated from disorder averaging serves to localize the corresponding dynamics.  

When the HSR model is combined with the disordered Anderson Hamiltonian the dynamics shows diffusive transport whenever $\Gamma>0$.\cite{moix2013,madhukar1977,rebentrost2008,kunsel2021} We examine the interplay between static disorder and dynamic noise with respect to transport further using the TT formalism. One can readily deduce that in the limit of an infinitesimal time step, $M(1)$ will be identical for all ensembles for which $\Gamma + \sigma^2\Delta t=CV$, with $C$ an arbitrary positive constant. Figure \ref{fig1} (b) shows the population as a function of time at the center site for a set of ensembles with $C=0.1$, in which all dynamics are initialized as a Kronecker delta. In one special case, an ensemble was prepared for $\sigma = V$ and $\Gamma=0$, but the resulting memory terms were artificially set to zero, i.e. $T(k)=0$ for all $k>1$ while keeping the first term the same. The resulting dynamics is identical to a $\sigma=0$, disorder-free and naturally memoryless propagation with $\Gamma=0.1V$. This demonstrates that when memory terms are eliminated, static and dynamic disorder cannot be differentiated.  In Fig. \ref{fig1} (b), several time steps are required before memory effects begin to manifest impacts on $M{(i)}$ and the populations from different ensembles diverge, with divergence occurring more quickly for larger $\sigma$. Regardless of the size of $\sigma$, as long as $\Gamma$ is nonzero all populations will eventually converge to the infinite temperature equilibrium state where all sites are equally occupied, indicated with a dashed black line in the figure. In the case in which $\sigma=V$ and there is no noise, the population on the site where the system was initialized remains above the equilibrium value.

We now examine select quantitative metrics that connect the tensors themselves to these dynamical characteristics. First, the magnitude of the tensors can serve as a direct measure of the significance of memory. We again quantify this with the tensor Frobenius norm. When there is finite noise, meaning there is diffusive transport and the system thermalizes, the tensor norms eventually decay to zero, indicating finite memory time. These finite memory times, which decrease with increasing $\Gamma$, are demonstrated in Fig. \ref{fig1} (c) alongside a $\Gamma=0$ case which shows eternal memory, as previously discussed.\cite{megier2017} 

Second, inspection of individual tensor elements can indicate transport properties. It is customary to follow the time-dependent spatial profile with certain localized initial conditions as in microscopy experiments to assess transport.\cite{swartzentruber1996,mensi2018,blach2025} To facilitate comparisons between different systems, averaged measures such as mean squared displacement $\langle x^2\rangle$, whose scaling with time can be used to classify the nature of transport, e.g. $\langle x^2\rangle\propto t$ for diffusive transport, are employed. Here, we take a different approach and define the outgoing-pseudoflux at time step $k$ for site $m$ as $f_m(k) = \sum_n' {T}_{n,n,m,m}(k),\;\; n \ne m$ and its cumulative sum over time $F_m(k) = \sum_{l=1}^{k} f_m(l)$. The former represents the magnitude of the influence that the population of the $m$th site $k$ steps in the past has on the populations of all other sites in the system at the current time step. The physical significance of $F_m(k)$ is further discussed in Appendix A, where we illustrate the association of $F_m(k>k_c)=0$ with localized systems for a certain finite $k_c$ value. Conversely, we associate $F_m(k)>0$ with delocalized systems. 

In Fig. \ref{fig1} (d) we examine $F_m(k)$ as $k$ increases. When $\Gamma>0$ and diffusion occurs, $F_m(k)$ converges to a positive number as $k \to \infty$ which, although not directly proportional to it, correlates with the non-monotonic behavior of the diffusion constant with $\Gamma$, behavior which is well understood.\cite{moix2013,rips1993,vcapek1993} Both the diffusion constant and $F_m(k)$ share a similar maximum location near $\Gamma=V$ for the chosen disorder strength. Note that, in the current model, the outgoing-pseudoflux $f_m(k)$ is always positive when $k=1$ and negative otherwise, indicating that the role of memory is to return the population back to its source, suppressing diffusion. This is not surprising given that removing memory is equivalent to introducing dynamical noise and destroying localization. We note that a different and perhaps more seamless connection between disorder and noise can be achieved with spatially uncorrelated but temporally correlated noise. Static disorder in this sense can be thought of as perfectly and indefinitely correlated noise while white noise is the opposite limit. In Appendix B we examine systems with exponentially correlated noise $\langle\epsilon_n(t)\epsilon_m(t')\rangle=\Gamma e^{-\gamma t}\delta_{nm}$, in which the two limits can be continuously bridged with a single parameter $\gamma$, the inverse correlation time. Similar conclusions on memory and transport can be drawn there.

\begin{figure}[h!]
\includegraphics[width=14cm]{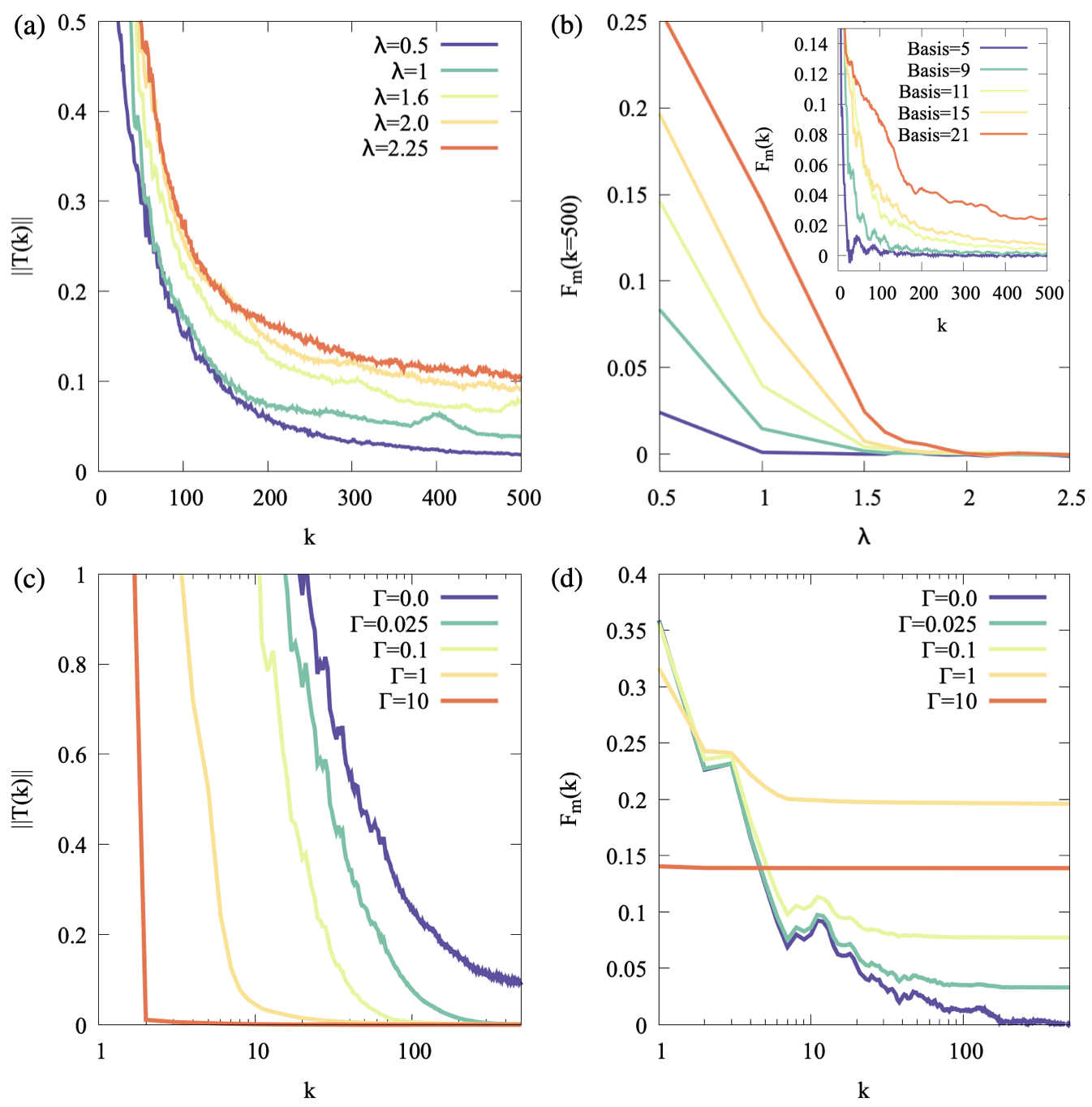}
\caption{ (a) The Frobenius norm of the transfer tensor $||T(k)||$ as a function of time step $k$ for several different $\lambda$ strengths.  (b) The cumulative outgoing-psuedoflux after $500$ time steps as a function of $\lambda$ for several different basis sizes of the AAH model, showing convergence towards the long-chain limit. The inset shows the convergence towards the final cumulative values as a function of time step for the varying basis sizes at a fixed $\lambda=1.5V$.  
(c) The Frobenius norm of the transfer tensor as a function of index given different $\Gamma$ for an AAH model with $\lambda=2V$. (d) The cumulative outgoing-pseudoflux as a function of time step for the same models as in (c).All calculations in this figure were run with $V\Delta t = 1/2$ and $2000$ trajectories and, unless otherwise specified, a basis of size $21$. Note that the norm of $T(1)$ is on the order of the basis size and not shown on the axes of (a) and (c).}
\centering
\label{fig2}
\end{figure}

\section{Transfer Tensor Analysis of Aubry-André-Harper Model}
In the above discussion, it might be tempting to conclude that eternal memory indicates localization. This misconception is easily debunked when one recalls the fact that a 3D Anderson model with a subcritical disorder strength could host localized and delocalized eigenstates simultaneously, separated by a mobility edge.\cite{grussbach1995,mott1995,semeghini2015} Since the source of eternal memory is independent of dimensionality, we conjecture that eternal memory derived from disorder averaging is a necessary but not sufficient condition for localization. 

To test this, we turn our attention to the Aubry-André-Harper (AAH) model, \cite{aubry1980,harper1955} which has been physically simulated with optical lattices.\cite{bar2007,roati2008,lahini2009,derrico2013,bordia2017,li2023} In the AAH model, instead of a random distribution, the site energies are given by $\epsilon_n = \lambda \cos(2\pi \beta n + \phi)$, where $\lambda$ is the strength of the potential, $\beta$ is an irrational number,  the golden ratio in our case, $\phi$ is a phase factor, and $-V$ replaces $V$ in Eq. \ref{eqham}. Here, the averaging is performed with respect to $\phi$, which is drawn from a uniform distribution spanning $[0,2\pi]$. The eigenstates are localized above a critical value of $\lambda=2V$ and delocalized below it.\cite{dominguez2019,aubry1980} We immediately see that, as shown in Fig. \ref{fig2} (a), this model displays eternal memory regardless of $\lambda$. Such memory does not provide an indication of localization, although a larger $\lambda$ results in convergence of the transfer tensor norm to a larger value.

At approximately $\lambda=2V$ in Fig. \ref{fig2} (b), there is a notable change in the behavior of $F_m(k)$, although the basis size has an influence on the localization to delocalization behavior. The cumulative outgoing-pseudoflux at the end of the simulation is shown as a function of $\lambda$ for several basis sizes. Smaller bases show less obvious delocalization to localization transitions, and these transitions appear at $\lambda<2V$. This is likely a result of density returning to its origin after reflecting off the system boundary. This leads to an underestimate of $\lambda$ for the localization transition in finite systems. Similarly, care must be taken in the Anderson model when $\sigma$ is so small that the eigenstate localization length exceeds the size of the finite basis. The largest basis size employed for the AAH model, $21$ in Fig. \ref{fig2} (b) is sufficiently converged towards the long-chain limit to clearly indicate the transition point at $\lambda=2V$. An inset in Fig. \ref{fig2} (b) shows how $F_m(k)$ converges to finite values as a function of $k$ for the case of $\lambda=1.5V$. 

HSR white noise can be introduced to the AAH model in precisely the same manner as in the Anderson model.\bibnote[noteone]{We note that there is a relevant but different type of dynamical noise in this case by adding a stochastic term $\sum_n\epsilon_n(t)|n\rangle\langle n|$ where $\epsilon_n(t)\propto\cos[2\pi\beta n+\phi_n(t)]$ with $\delta$-correlated phase factor $\phi_n(t)$. Upon averaging this leads to the same first TT as in the normal disordered AAH model in analogy to the relationship between the HSR and Anderson model. However, this type of noise produces dynamics that is quite involved and will be left for a future investigation. } Inclusion of heat baths in this model has been previously reported to induce transport.\cite{bhakuni2024} Introduction of noise results in a finite memory time, as demonstrated in Fig. \ref{fig2} (c). The AAH model also shows the same kind of non-monotonic behavior of $F_m(k)$ with increasing $\Gamma$ as seen in the Anderson model, with the outgoing-pseudoflux converging to non-zero values which are maximized at a moderate value of $\Gamma \approx V$, suggesting non-monotonic behavior of transport efficiency with noise. 

As the AAH model has been realized many times, this phenomenon could be experimentally confirmed provided that the noise strength could be controlled sufficiently.\cite{bar2007,roati2008,lahini2009,derrico2013,bordia2017,li2023}

\section{Summary}
Applying the transfer tensor methods to simple models of transport in lattices has demonstrated how memory effects differentiate static and dynamic disorder while employing much smaller bases than are needed for conventional numerical studies of transport properties. The structure of the tensors explains the meaning of the eternal memory which arises when performing an ensemble average over static disorder and restoring spatial homogeneity. Memory removes fictitious terms that would correspond to redrawing static disorder thereby converting it to dynamic disorder. This eternal memory, while necessary for localization, is not a sure indication of it. However, localization can be inferred from the total outgoing-pseudoflux. The outgoing-pseudoflux behaves similarly to the diffusion constant in the Anderson-HSR model, perhaps indicating that more complicated metrics of the tensors might produce the diffusion constant itself or indicate whether a system is diffusive or ballistic. Future inquiries in this direction are warranted as transfer tensor methods show promise as an analysis tool for understanding memory effects and addressing transport properties.

\section*{Acknowledgment} 
We thank the University of Nevada, Las Vegas for supporting this research.

\section*{Data Availability}
The data and code to support this study are available from the corresponding author on reasonable request.

\section*{Author Contributions}
Michelle C. Anderson and Chern Chuang contributed equally to this work.

\section*{Appendix A: Steady-state analysis of transfer tensors for localized systems} \label{sec:SS} 
Before we examine the steady state condition, it is useful to review some mathematical properties of TT in the current context. We assume that a set of TTs, $T_{mm'nn'}(l)$ with $l=1,2,\cdots$, is obtained by the homogenization procedure described in the main text of a spatially disordered lattice model. In addition to the translational invariance symmetry discussed in the main text, we have
\begin{eqnarray}
    T_{mm'nn'}(l)&=&T_{nn'mm'}(l)\label{eqn:timereversal}\\
    \sum_{m}T_{mmnn}(l)&=&\delta_{l1}\label{eqn:dmproperty}
\end{eqnarray}
where Eq.~(\ref{eqn:timereversal}) follows from time-reversal symmetry. Eq.~(\ref{eqn:dmproperty}) can be understood as the first tensor being a legitimate dynamical map and, consequently, norm-conserving, while the rest of the tensors do not add to the populations. \\

For a finite, delocalized system subject to the HSR infinite temperature bath, we expect the steady-state solution to be $\rho^{(ss)}_{nn}=p_n^{(ss)}=1/N$, where $p_n^{(ss)}$ is the steady state population of site $n$, regardless of the initial state.\bibnote[note2]{The treatise for coherence will be left for an upcoming report. Here, we focus on the determination of localization in the steady state and population suffices.} On the other hand, for localized systems, we anticipate an initial state-dependent steady state solution that is generally spatially inhomogeneous, $\rho_{nn}^{(ss)}\neq\rho_{mm}^{(ss)}$. By substituting $\rho^{(ss)}$ into Eq.~(\ref{Eq:tt_prop}), we have
\begin{eqnarray}
    \rho^{(ss)}=\sum_{l=1}^kT(l)\rho^{(ss)}=\mathbb{T}(k)\rho^{(ss)}\label{eqn:ss}
\end{eqnarray}
where $\mathbb{T}(k)$ is the cumulative sum of the tensors. Theoretically Eq.~(\ref{eqn:ss}) holds only in the limit of $k\rightarrow\infty$ due to eternal memory. However, by applying Eqs.~(\ref{eqn:timereversal}) and (\ref{eqn:dmproperty}) as well as translational symmetry we get
\begin{eqnarray}
    \left[\bar{T}(k)-\mathbb{I}\right]\cdot\vec{p}^{(ss)}=\begin{bmatrix}
        \bar{T}_0(k)-1 & \bar{T}_1(k) & \cdots & \bar{T}_{N-1}(k)\\
\bar{T}_1(k) & \bar{T}_0(k)-1 & \cdots & \bar{T}_{N-2}(k)\\
\bar{T}_2(k) & \bar{T}_1(k) & \cdots & \bar{T}_{N-3}(k)\\
\vdots&\vdots&\vdots&\vdots\\
\bar{T}_{N-1}(k) & \bar{T}_{N-2}(k) & \cdots & \bar{T}_0(k)-1
    \end{bmatrix}\begin{bmatrix}
p_{1}^{(ss)}\\p_{2}^{(ss)}\\\vdots\\p_{N}^{(ss)}
\end{bmatrix}=0
\end{eqnarray}
where $\bar{T}_\Delta(k)=\mathbb{T}_{00\Delta\Delta}(k)$ and $\sum_\Delta \bar{T}_\Delta=1$. 

Now, since $\vec{p}^{(ss)}$ is generally a non-uniform vector and changes upon changing the initial condition for a localized system, the only possible solution to this overdetermined set of equations is $\bar{T}=\mathbb{I}$. In practice, we find this condition to hold for $\bar{T}(k\ge k_c)$ for a finite $k_c$, at which $F_m(k\ge k_c)=0$ also holds as discussed in the main text. We also note that for an arbitrary initial condition, it may take more time steps than $k_c$ to reach the steady state. When the system is not localized, the above argument does not hold and we cannot restrict the form of $\bar{T}$ in this way. Numerically, we observe $F_m(k)>0$ for delocalized systems.

The current framework provides limited insight into the numerical correspondence of the outgoing-pseudoflux and the diffusion constant observed in the Anderson-HSR model, while the correlation between the two are strongly implicated by their similar non-monotonic behavior with respect to $\Gamma$ as shown in Fig.~\ref{fig2}. Some intuition can be developed from the Green-Kubo expression for the diffusion constant
\begin{eqnarray}
    D=\frac{1}{Z}\int_0^\infty dt~\mathrm{Tr}[e^{-\beta H}\hat{j}(t)\hat{j}(0)]\label{eqn:GreenKubo}
\end{eqnarray}
where $\hat{j}(t)$ is the flux operator and $e^{-\beta H}/Z=N^{-1}$ is the partition function for the infinite-temperature HSR model.\cite{Chuang2016} The time-evolution of operators in the Heisenberg picture can be similarly defined as in Eq.~(\ref{Eq:tt_prop}) albeit with a complex conjugate. In the current model of nearest-neighbor tight-binding chain, we have $\hat{j}=iV\sum_n|n\rangle\langle n+1|+H.c.$.  With this, it becomes clear that the integrand in Eq.~(\ref{eqn:GreenKubo}) is in fact related to a quantity different from $F_m(k\rightarrow\infty)$:
\begin{eqnarray}
    \frac{1}{N}\mathrm{Tr}[\hat{j}(k\Delta t)\hat{j}(0)]=2V^2\sum_n\left[M_{n+1,n,1,0}(k)-M_{n-1,n,1,0}(k)\right].
\end{eqnarray}
And the diffusion constant can be approximated by the $k$-summation of this quantity in the $\Delta t\ll1$ limit. Note that disorder-averaged dynamical maps $M(k)$'s are used, instead of TT. This is indicative of the physical significance and the distinction between the maps and TT. To track the temporal evolution of a dynamical quantity (flux in this case), one needs to account for all possible pathways leading to the current time step, Markovian or not, signified by the nested tensor multiplications in Eq.~(\ref{Eq:tt_prop}).

\begin{figure}[h!]
\includegraphics[width=14cm]{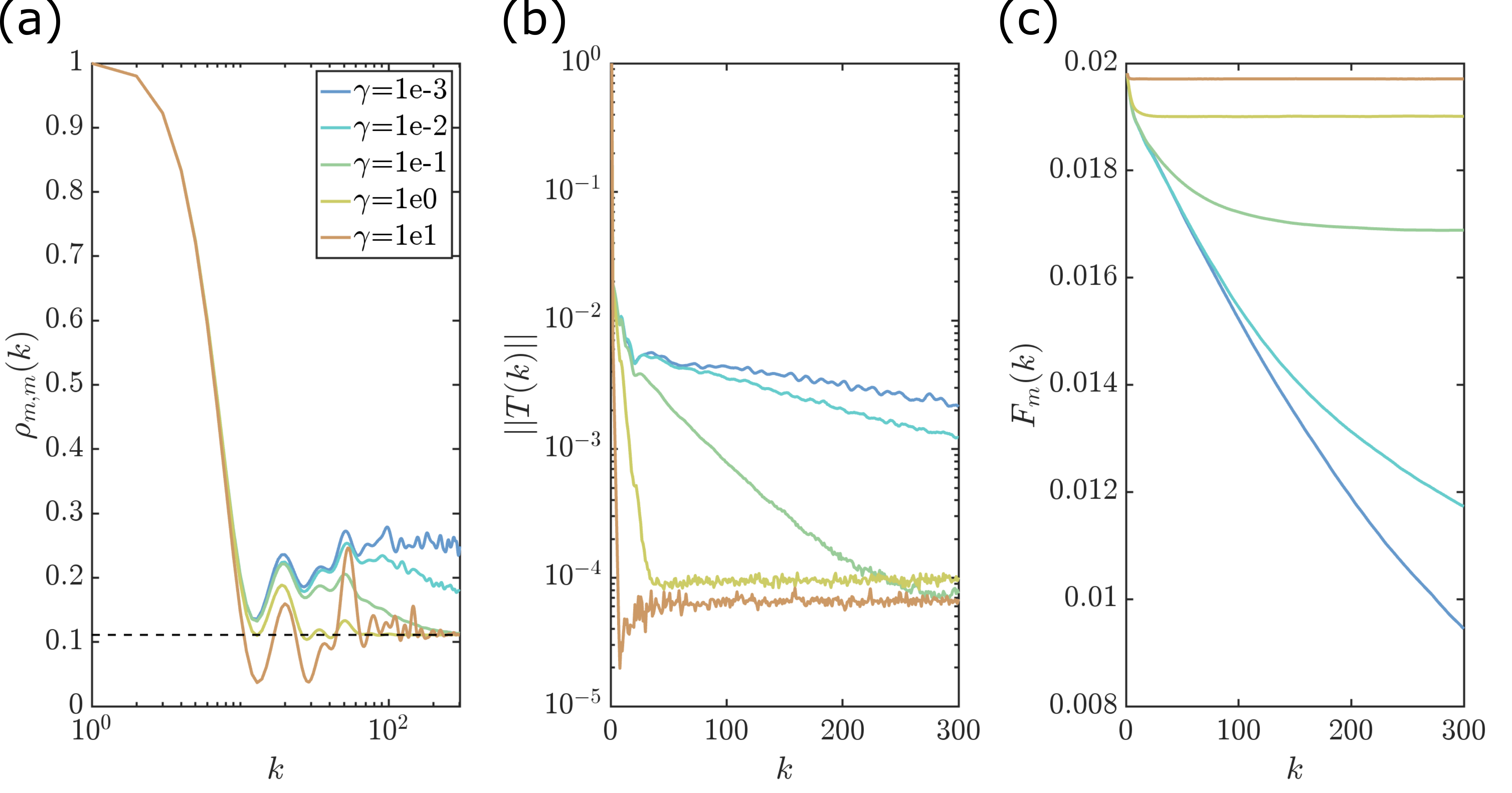}
\caption{Systems with exponentially correlated noise ($\langle\epsilon_n(t)\epsilon_m(t')\rangle=\Gamma e^{-\gamma |t-t'|}\delta_{nm}$) with 9 sites, $V\Delta t=0.1$, $\Gamma/V=1$, and various $\gamma$ values. (a) Population at the central site as a function of time step and $\gamma$. The full thermalization limit $\rho_{m,m}(t\rightarrow\infty)=N^{-1}$ is indicated by the dashed line. (b) Tensor Frobenius norm. The plateauing for smaller $\gamma$ curves suggest more noise samples are needed for convergence. With more noise samples the plateau continues to decrease in value. (c) Outgoing-pseudoflux. }
\centering
\label{figA1}
\end{figure}

\section*{Appendix B: Correlated Noise} \label{sec:CorrNoise} 
The interplay between static disorder and dynamic noise manifested in the mixed Anderson-HSR model can be analogously present in systems with correlated noise. Disorder can be thought of as perfectly correlated, constant noise, while the HSR white noise is the opposite limit with vanishing temporal correlation. Here, we demonstrate this analogy using a system with temporally exponentially correlated (but spatially independent) on-site energies, i.e. $\langle \epsilon_n(t)\epsilon_m(0)\rangle=\Gamma e^{-\gamma t}\delta_{nm}$. With this model, two free parameters $\Gamma$, the noise strength, and $\gamma$, the inverse correlation time exist. In the limits of $\gamma\rightarrow\infty$ and $\gamma\rightarrow0$ one recovers the HSR and the Anderson models, respectively. 

The numerical results are shown in Fig. \ref{figA1} in the same convention as in Fig.~\ref{fig1}. As predicted, the dynamics and the corresponding tensors follow the same trend as in the mixed Anderson-HSR models studied in the main text. In Fig. ~\ref{figA1} (a) we examine the convergence of the dynamics and the TT in the short time limit for systems with same noise strength $\Gamma$ and varying correlation time $\gamma^{-1}$. In part (b) the tensor norms, after a certain transient regime around $k=30$ for larger $\gamma$ values, decay single exponentially in accordance with the rates $\gamma$. Finally, we show the outgoing-pseudoflux in part (c). In contrast to the mixed Anderson-HSR models with constant disorder and increasing noise strength shown in Fig. ~\ref{fig1} (d), here the pseudoflux increases monotonically with increasing $\gamma$.

\section*{References}
\bibliographystyle{unsrt}
\bibliography{references}

\end{document}